\documentclass[twoside]{dis07}
\usepackage[latin1]{inputenc}
\usepackage[dvips]{graphicx,epsfig,color}
\usepackage{wrapfig,rotating}
\usepackage{amssymb,amsmath,array}

\begin{document}
\title{New Global Fit to the Total Photon-Proton Cross-Section $\sigma_{L+T}$ 
and to the Structure Function $F_2$}

\author{Dominik Gabbert$^1$ and Lara De Nardo$^{1,2}$
\vspace{.3cm}\\
1- DESY, 22603 Hamburg, Germany
\vspace{.1cm}\\
2-TRIUMF, Vancouver, British Columbia V6T 2A3, Canada \\
}

\maketitle

\begin{abstract}
A fit to world data on the photon-proton cross section
$\sigma_{L+T}$ and the unpolarised structure function $F_2$ is
presented. The 23-parameter ALLM model based on Reggeon and Pomeron exchange is
used. Cross section data were reconstructed to avoid inconsistencies with respect to $R$ of the published $F_2$ data base. Parameter uncertainties and correlations are obtained.
\end{abstract}

\section{Introduction}
Deep-inelastic scattering on protons has been studied precisely in the last
decades at various energies covering a large kinematic region provided by collider and fixed target experiments, thus providing us with our modern understanding of the  proton structure.

The inclusive DIS cross section in the one-photon-exchange approximation 
 is related to the unpolarized structure function $F_2(x,Q^2)$ and the
ratio $R(x,Q^2)$ of longitudinal and transverse photo-absoption cross section:
\begin{equation}\label{sigma}
\frac{d^2\sigma}{dx~dQ^2}=
\frac{4\pi\alpha^2_{em}}{Q^4}\frac{F_2(x,Q^2)}{x}\left\{1-y-\frac{Q^2}{4E^2}+
\left(1-\frac{2m^2}{Q^2}\right)\frac{y^2+Q^2/E^2}{2[1+R(x,Q^2)]}\right\}~.
\end{equation}
Here, $Q^2$ is the square of the photon 4-momentum and $x=Q^2/2M\nu$ with the proton mass M and the photon energy $\nu$ in the proton rest frame.

From Eq.~(\ref{sigma}) it follows that a
measurement of the cross section alone is not
sufficient to extract both, $F_2$ and $R$, and that only a
variation of the beam energy $E$ in the proton rest frame for fixed kinematic conditions can give access 
to both quantities. 
Alternatively, $F_2$ can be extracted using parameterizations of world data on $R$:
two common examples are $R_{1990}$~\cite{r1990} and
$R_{1998}$\cite{r1998}, whose differences reflect the states of world
knowledge at the time they were obtained. 
The sensitivity of the cross section to $R$ increases with $y$ 
as it can be seen in Eq.~(\ref{sigma}). 
The discrepancy in the extracted values of $F_2$ using the two parameterizations can  exceed  4\% in the regions of maximum $y$. 

The structure function $F_2$ is related to the photon-proton 
cross section $\sigma_{L+T}$ by the expression: 
\begin{equation}
\sigma_{L+T}=\frac{4\pi^2\alpha_{em}}{Q^4}\frac{Q^2+4M^2x^2}{1-x}F_2~.
\end{equation}

For virtual photons this relation employs the Hand convention for the
virtual photon flux. It was used for technical convenience of
consistency between real and virtual photon processes. 

This paper reports on a new fit of the photon-proton cross section
$\sigma_{L+T}$  which reflects the recent world knowlege on the cross
section and is self-consistent with respect to the use of $R$, since
the cross sections were reconstructed in each case using the value of
$R$ that had been used to extract the published values of $F_2$. A
result of the fit is a facility to calculate values of $F_2$ based on
a single parameterization of $R=R_{1998}$. 

\section{The fit }
The fit includes 2740  data points: 
 574 from the SLAC  experiments E49a, E49b, E61, E87, E89a, E89b~\cite{slac};
292  from NMC~\cite{nmc}; 787 from H1~\cite{h1}; 570 from 
ZEUS~\cite{zeus};
91  from E665~\cite{e665}; 229 points from BCDMS~\cite{bcdms}. Real
photon data comprise 196 points 
from Ref.~\cite{pdgreal} and 1 from ZEUS~\cite{zeusreal94}. 

\begin{wraptable}{r}{0.614\columnwidth}
\centerline{\begin{tabular*}{0.55\columnwidth}[h]{|c|c|c|c|}
\hline
Parameter &  ALLM97  & this fit & uncertainty   \\
\hline
$m_0^2($GeV$^2)$            & 0.31985 & 0.454 & 0.137   \\
$m_\mathcal{P}^2($GeV$^2)$  & 49.457  & 30.7 & 13.4   \\
$m_\mathcal{R}^2($GeV$^2)$  & 0.15052 & 0.118 & 0.224   \\
$Q_0^2($GeV$^2)$            & 0.52544 & 1.13 & 1.47   \\
$\Lambda_0^2($GeV$^2)$      & 0.06527 & 0.06527 & -   \\
$a_{{\mathcal{P}}1}$       & -0.0808 & -0.105 & 0.024   \\
$a_{{\mathcal{P}}2}$       & 0.44812 & -0.496 & 0.154   \\
$a_{{\mathcal{P}}3}$       & 1.1709  & 1.31   & 1.04    \\
$b_{{\mathcal{P}}4}$       & 0.36292 & -1.43 & 2.31   \\
$b_{{\mathcal{P}}5}$       & 1.8917  & 4.50  & 2.46   \\
$b_{{\mathcal{P}}6}$       & 1.8439 & 0.554  & 0.531  \\
$c_{{\mathcal{P}}7}$       & 0.28067 & 0.339 & 0.093  \\
$c_{{\mathcal{P}}8}$       & 0.22291 & 0.128 & 0.104  \\
$c_{{\mathcal{P}}9}$       & 2.1979 & 1.17 & 1.14   \\
$a_{{\mathcal{R}}1}$       & 0.584  & 0.373 & 0.150   \\
$a_{{\mathcal{R}}2}$       & 0.37888& 0.994 & 0.443   \\
$a_{{\mathcal{R}}3}$       & 2.6063 & 0.781 & 0.524   \\
$b_{{\mathcal{R}}4}$       & 0.01147& 2.70 & 1.84   \\
$b_{{\mathcal{R}}5}$       & 3.7582 & 1.83 & 2.39   \\
$b_{{\mathcal{R}}6}$       & 0.49338& 1.26 & 1.33   \\
$c_{{\mathcal{R}}7}$       & 0.80107& 0.837& 0.500  \\
$c_{{\mathcal{R}}8}$       & 0.97307& 2.34 & 2.34   \\
$c_{{\mathcal{R}}9}$       & 3.4942 & 1.79 & 0.93   \\
\hline
\end{tabular*}}
\caption{Parameters of the functional form used in the ALLM parameterization ~\cite{allm91}. Results of the ALLM97 fit ~\cite{allm97} without uncertainties in comparison to the results discussed in this paper with uncertainties. 
These uncertainties correspond only to the diagonal elements of the full covariance matrix which must be used to calculate uncertainties in $F_2$ or cross sections. The parameter $\Lambda^2_0$ has no uncertainty as it was fixed in the fit.}
\label{tab:parameters}
\end{wraptable}

The ALLM functional form is a 23-parameter model 
of $\sigma_{L+T}$ where $F_2$ is described by  Reggeon and Pomeron
exchange, valid for $W^2>4\,$GeV$^2$, i.e., above
the resonance region, and any $Q^2$ including the real $\gamma$ process. Here, $W^2$ is the invariant squared mass of the photon-proton system.
For details on the  parameterization we refer to the original
papers ~\cite{allm91,allm97}. 
The new fit was performed by minimizing the $\chi^2$ defined in Eq.~(\ref{eq:chi2}) where  $D_{i,k}\pm \sigma_{i,k}^{stat}\pm\sigma_{i,k}^{syst}$ are
the values of $\sigma_{L+T}$ for 
data point $i$ within the data set $k$, 
$\delta_k$ is the normalization uncertainty in data set $k$ quoted by the
experiment, $\nu_k$ is a parameter for
the normalization of each data set in  units of the normalization
uncertainty, $T({\bf p},W^2,Q^2)$ is the functional form of the 23-parameter ALLM parameterization.

The $\chi^2$ takes into account uncorrelated point-by-point
statistical and systematic 
uncertainties and overall normalization uncertainties. 
The normalization parameters $\nu_k$  determine
 the size of the shifts in units of the normalization uncertainties
 $\delta_k$.

\begin{eqnarray}
\label{eq:chi2}
\nonumber \chi^2({\bf p},{\mbox{\boldmath $\nu$}}) =
\sum_{i,k}{\frac{[D_{i,k}(W^2,Q^2)\cdot(1+\delta_k{\mathbf \nu_k})
    -T({\bf{p}},W^2,Q^2)]^2 
  }{({\sigma_{i,k}^{stat}}^2+{\sigma_{i,k}^{syst}}^2)
\cdot(1+\delta_k\nu_k)^2}}+ 
\sum_k{\nu_k^2}
\\ 
 \approx
 \sum_{i,k}{\frac{ [ D_{i,k}(W^2,Q^2) - T({\bf{p}},W^2,Q^2)  \cdot
     (1-\delta_k\nu_k) 
     ]^2 }
{{\sigma_{i,k}^{stat}}^2+{\sigma_{i,k}^{syst}}^2}}+\sum_k{\nu_k^2} ~,
\end{eqnarray}

In order to keep the number of free parameters as small as possible, the
normalization parameters are determined analytically 
in each minimization step using the relation
\begin{equation}
\nu_k
=\frac{\sum_i{\delta_kT_{i,k}(T_{i,k}-D_{i,k})/\sigma_{i,k}^2}}{
\sum_iT_{i,k}^2\delta_k^2/\sigma_{i,k}^2+1},
\end{equation}
obtained by requiring $\partial \chi^2/\partial \nu_k=0$ in the context of the approximation for $\chi^2$ in the second line of
Eq.~(\ref{eq:chi2}); here
$\sigma_{i,k}^2={\sigma_{i,k}^{stat}}^2+{\sigma_{i,k}^{syst}}^2$.  
This separate extraction is possible since the normalization
parameters are not correlated and depend only on the involved data points and
the functional parameters.
The resulting fit has a reduced $\chi^2$ equal to 0.94; the
contributions from each data set, together with the normalization
parameters can be found in Ref.~\cite{dis07talk}.
Table~\ref{tab:parameters} shows the final parameters from this fit
with the corresponding uncertainties and, for comparison, the
parameters from the ALLM97 fit. Figure~\ref{Fig:sigmatot} shows the
new fit in comparison with world data and with the ALLM97 fit. 
A full comparison between the two fits
is not possible as in the ALLM97 fit parameter uncertainties were not
provided.
Presumely, these uncertainties are larger than the those of the new fit, since the size of the current data set is nearly twice as large.
The uncertainties in the cross sections calculated from 
the fit as represented by the error bands in the figure
are much smaller than individual error bars on the original
data points because of the smoothness constraint inherent
in the fitted model.  The fit evaluated at any kinematic
point is effectively an average of a number of data points.

In conclusion, a new fit of world data on $\sigma_{L+T}$ and $F_2$ is
presented. Such a fit is consistent in the choice of the $R$
parameterization $R_{1998}$.  
Also, for the first time, parameter and fit uncertainties are calculated.
A subroutine that allows the calculation of $\sigma_{L+T}$ and $F_2$
with their fit uncertainties is available upon request from the
authors.

\section{Acknowledgments}

The authors would like to thank E.C.~Aschenauer, A.~Miller and
W.-D.~Nowak for valuable discussions.

\begin{footnotesize}


\end{footnotesize}


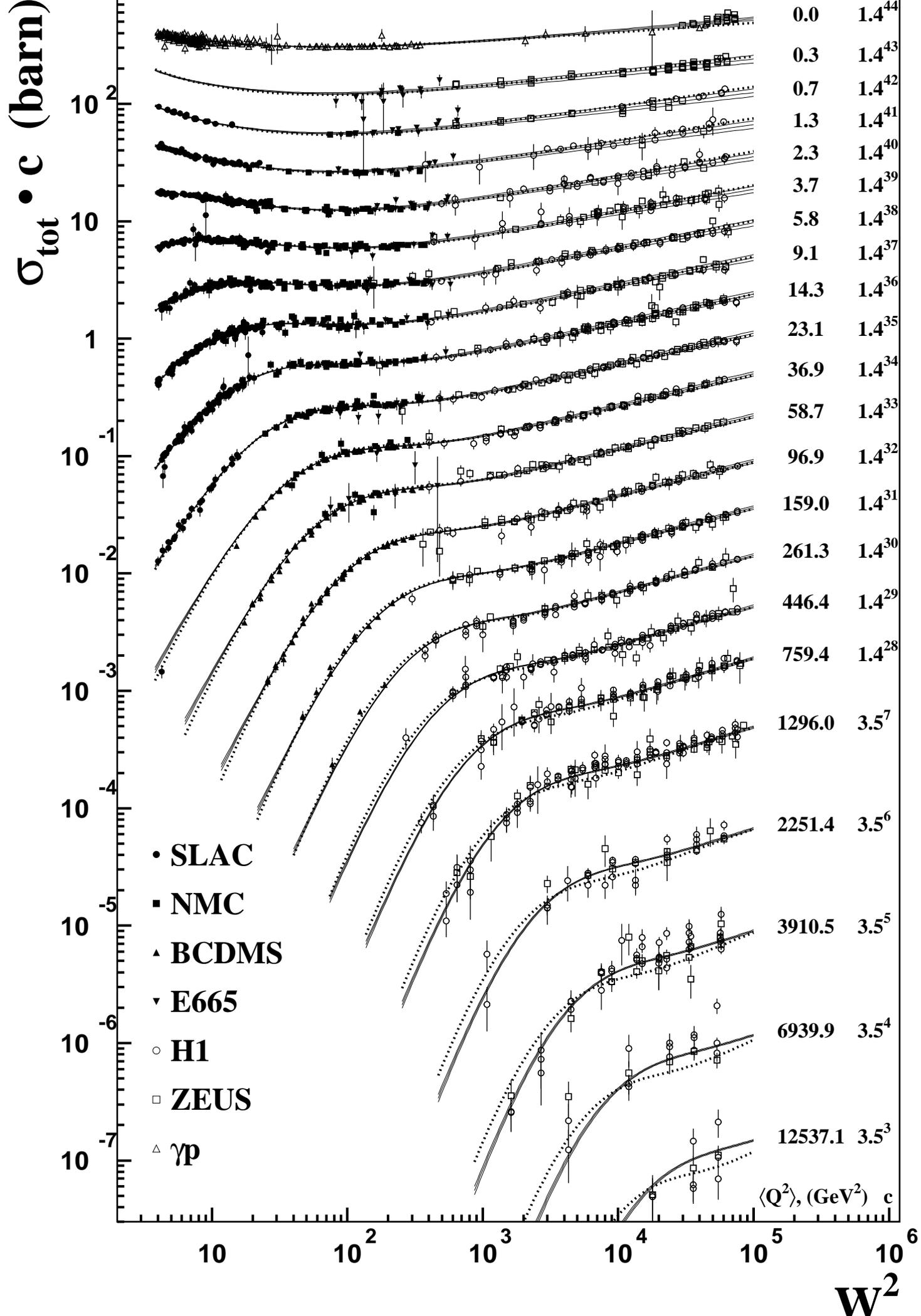
\end{document}